\documentclass[12pt,a4paper,superscriptaddress,showpacs,preprint]{revtex4-1}
\usepackage[dvips,dvipdfm]{graphicx,color}
\usepackage{graphicx}
\usepackage{latexsym}
\usepackage{amsmath,amssymb}

\newcommand{\ie}{\textit{i.e.}}
\newcommand{\Eif}{E_{{\rm if}}}
\newcommand{\Ei}{E_{{\rm i}}}
\newcommand{\Me}{M_{{\rm e}}}

\newcommand{\taue}{\tau_{{\rm e}}}
\newcommand{\taui}{\tau_{{\rm i}}}

\newcommand{\EQ}{Eq.~}
\newcommand{\EQS}{Eqs.~}
\newcommand{\FIG}{Fig.~}
\newcommand{\FIGS}{Figs.~}

\newcommand{\SEC}{Sec.~}

\begin{document}

\title{Voter model with non-Poissonian interevent intervals}

\author{Taro Takaguchi}
\affiliation{Department of Mathematical Informatics, The University of Tokyo, 7-3-1 Hongo, Bunkyo, Tokyo 113-8656, Japan}
\author{Naoki Masuda}
\affiliation{Department of Mathematical Informatics, The University of Tokyo, 7-3-1 Hongo, Bunkyo, Tokyo 113-8656, Japan}
\affiliation{PRESTO, Japan Science and Technology Agency, 4-1-8 Honcho, Kawaguchi, Saitama 332-0012, Japan}

\begin{abstract}
Recent analysis of social communications among humans has revealed that the interval between interactions for a pair of individuals and for an individual often follows a long-tail distribution. We investigate the effect of such a non-Poissonian nature of human behavior on dynamics of opinion formation. We use a variant of the voter model and numerically compare the time to consensus of all the voters with different distributions of interevent intervals and different networks. Compared with the exponential distribution of interevent intervals (\ie, the standard voter model), the power-law distribution of interevent intervals slows down consensus on the ring. This is because of the memory effect; in the power-law case, the expected time until the next update event on a link is large if the link has not had an update event for a long time. On the complete graph, the consensus time in the power-law case is close to that in the exponential case. Regular graphs bridge these two results such that the slowing down of the consensus in the power-law case as compared to the exponential case is less pronounced as the degree increases.
\end{abstract}


\maketitle

\section{Introduction}
Macroscopic social dynamics often occur as a result of microscopic dynamics of individuals
interacting on networks of social contacts.
Studies of interacting particle systems such as spin systems
have enriched our understanding of various types of social dynamics
such as epidemics, information cascades, opinion formation,
synchronization, and evolutionary games~\cite{Barrat2008dpo_book,Castellano2009spo,Newman2010nai_book,Easley2010nca_book}.
It is established that the structure of
social networks influences social dynamics in many different ways.
 
Social dynamics on networks are often modeled by stochastic processes.
Models of these dynamics usually assume
that interaction events between a pair of individuals occur according to the Poisson process.
This assumption facilitates theoretical analysis of models
and corresponds to the situation where the rate at which an event occurs
generally depends on the current configuration of the network
but not on the history of the dynamics.

Recent developments of sensor technologies and accumulation of massive amounts of electronic data
have facilitated detailed analysis of point processes related to social dynamics of humans. 
Examples of such data include email exchanges~\cite{Eckmann2004eod,Barabasi2005too,AVazquez2006mba,Malmgren2008ape,Rybski2009slo},
cell-phone calls~\cite{Candia2008uia,Song2010lop},
and face-to-face conversations~\cite{Barrat2008hrd,Isella2010wia,Cattuto2010dop}. 
Apart from the network structure, 
these studies have revealed a novel universal feature of social dynamics:
the non-Poissonian interevent intervals (IEIs).
The distributions of IEIs are often inherited with long tails
and can be modeled by the power-law distribution possibly with an exponential cutoff~\cite{AVazquez2006mba,AVazquez2007ion}
or by the log-normal distribution~\cite{Iribarren2009ioh}.
The long-tail IEI distribution can be explained by the prioritization of tasks~\cite{Barabasi2005too,AVazquez2006mba}
or by the combination of seasonality and the circadian rhythm of human activities~\cite{Malmgren2008ape,Hidalgo2006cft}.

The effect of the long-tail IEI distributions on epidemic dynamics 
such as the susceptible-infected (SI) and susceptible-infected-recovered (SIR) models
has been investigated~\cite{Barrat2008hrd,Isella2010wia,AVazquez2007ion,Iribarren2009ioh,Karsai2010sbs,Min2010sdf,Karrer2010mpa}. 
It has been suggested that the long-tail IEI distribution is responsible for 
the persistent prevalence of computer viruses~\cite{AVazquez2007ion}
and email advertisements~\cite{Iribarren2009ioh}
that cannot be explained by the Poisson assumption.
The slowing down of epidemic dynamics owing to the long-tail IEI distributions
is also found in the epidemic model on priority-queue networks~\cite{Min2010sdf}.

In this paper, we consider the effect of the long-tail IEI distribution on opinion dynamics. 
We use a variant of the voter model and compare the time required by this variant to achieve consensus among all the voters
with that required by the standard voter model with the exponential IEI distribution.
By numerical simulations,
we show that the long-tail IEI distribution increases the consensus time  on the ring.
The consensus time obtained for the exponential and long-tail IEI distributions is relatively close on the complete graph.
We interpolate the results on the ring and the complete graph by examining the voter models
on the regular random graph;
by this interpolation, we show that the node degree is a main determinant of the difference in the consensus time.

\section{Model}\label{sec:model}

We analyze a variant of the voter model~\cite{Castellano2009spo,Clifford1973amf,Holley1975etf,Redner2001agt_book}
on static regular networks; \ie, all the nodes have the same degree.
A voter is placed at each node in the network, and each voter takes one of the two opinions
denoted by {\bf 0} and {\bf 1}.
Initially, each voter takes {\bf 0} and {\bf 1} with equal probability (\ie, $0.5$),
and the opinions of different voters are assigned independently.
Each link between the nodes is independently endowed with a random IEI $\tau_1$,
which represents the time until the initial update event occurs on this link.
We denote the distribution of $\tau_1$ by $p_1(\tau_1)$.
Suppose that an update event on a link occurs at a certain time.
If the two endpoints of the link are occupied by the opposite opinions,
one of the two voters is selected with equal probability (\ie, 0.5)
and the opinion of the selected voter is flipped such that the two voters take the same opinion.
Otherwise, nothing happens in the update event.
Then, the next IEI for this link is drawn from distribution $p(\tau)$.
The sequence of update events on each link is a renewal process~\cite{Cox1967rt}
and only the initial IEI $\tau_1$ obeys $p_1(\tau_1)$, whereas all the subsequent IEIs obey $p(\tau)$.
If $\tau_1$ were taken from distribution $p(\tau)$, we would be implicitly assuming that there is an event at time 0. Introduction of $p_1(\tau_1)$ is necessary to avoid such artificial initial conditions.
Update events and the assignment of random IEIs occur independently on all the links.
It should be noted that update events never (\ie, with measure zero) occur on multiple links at the same time
if $p_1(\tau_1)$ and $p(\tau)$ do not possess point mass, which we assume.
We assume that each node is updated once per unit time such that $\delta t = 1/N$.

In this study, we assume that $p_1(\tau_1)$ obeys
\begin{equation}
p_1(\tau_1) =  \frac{1}{\langle \tau \rangle}\int_{\tau_1}^\infty p(\tau^\prime) d\tau^\prime,
\label{eq:p1}
\end{equation}
where $\langle \cdot \rangle$ denotes the mean.
With the definition of $p_1(\tau_1)$, the renewal process is the so-called equilibrium renewal process~\cite{Cox1967rt}.
If the renewal process is ongoing and the voter dynamics begin at an arbitrary instant,
then $\tau_1$ obeys $p_1(\tau_1)$ given by \EQ\eqref{eq:p1}, not $p(\tau)$.

We run the dynamics until the entire network is taken over by one opinion;
we refer to such a unanimous configuration as a consensus.
We are concerned with the consensus time,
\ie, the time required to reach a consensus.
To obtain the averaged quantities,
we perform 1000 rounds of simulations under each condition unless otherwise stated.

In the present study,
we examine the following two voter models.
\begin{itemize}
\item For the exponential voter model, we set
\begin{eqnarray}
p(\tau) &=& \lambda \exp(-\lambda \tau),\\
p_1(\tau_1) &=& \lambda \exp(-\lambda \tau_1).
\end{eqnarray}
\item For the power-law voter model, we set
\begin{eqnarray}
p(\tau) &=&\frac{\alpha-1}{c} \left( \frac{\tau+c}{c} \right)^{-\alpha}, \label{eq:p_power}\\
p_1(\tau_1) &=&  \frac{c}{\alpha-2} \left( \frac{\tau_1 +c}{c} \right)^{-(\alpha-1)}. \label{eq:p1_power}
\end{eqnarray}
\end{itemize}
In the exponential voter model, the occurrence of events on each link obeys the independent and identical Poisson process.
The exponential voter model is equivalent to the standard voter model
with the updating procedure called link update~\cite{Suchecki2005clf} or link dynamics (LD)~\cite{Sood2008vmo}.

For the power-law voter model,
our choice of the power-law $p(\tau)$ is motivated by recent experimental results
obtained for face-to-face interactions~\cite{Barrat2008hrd}.
We assume $\alpha > 2$ such that the power-law $p(\tau)$ has a finite mean.
Empirically, $p(\tau)$ usually has a value of $\alpha$ less than two~\cite{Eckmann2004eod,Barabasi2005too,AVazquez2006mba,Malmgren2008ape,Candia2008uia,Song2010lop,Barrat2008hrd,AVazquez2007ion}
and accompanies an exponential cutoff~\cite{AVazquez2006mba,AVazquez2007ion,Candia2008uia}.
With $\alpha \leq 2$,
the consensus time of the power-law voter model trivially diverges for any network
because $\langle \tau\rangle$ diverges.
Instead of setting $\alpha<2$ and modulate $\alpha$ or the cutoff value of $\tau$,
we use \EQ\eqref{eq:p_power} with different values of $\alpha > 2$ to examine the effect of the large variance
of interevent intervals on the consensus time.
  
We set $\lambda = 1$ and $c=\alpha-2$ 
to set $\langle \tau \rangle$ to unity for both the exponential and power-law voter models.
In our numerical simulations, we set $(\alpha, c) = (2.5,0.5)$ and $(3.5,1.5)$.
We verified that our numerical results presented in the following sections for $\alpha=2.5$
are qualitatively the same in the range $2 < \alpha < 3$
and that the results for  $\alpha=3.5$ are qualitatively the same in the range $3 < \alpha < 4$.

The heterogeneous degree distribution of the network,
which is eminent in scale-free networks, makes the consensus time
sensitive to the adopted update rule~\cite{Suchecki2005clf,Sood2008vmo}.
The consensus time in networks with heterogeneous degree distribution
has been investigated for the exponential (\ie, standard) voter model~\cite{Suchecki2005clf,Sood2008vmo,Sood2005vmo,Castellano2005cov,Suchecki2005vmd,Antal2006edo,FVazquez2008aso}.
To focus on the effect of the power-law IEI distribution, 
we restrict ourselves to the regular graphs in the present study.
Specifically, we compare the consensus time of the exponential and power-law voter models
on the ring, the complete graph, the extended ring, and the regular random graph.

\section{Results}
\subsection{Ring}\label{sec:ring}
Assume that the voters are placed on the ring with $N$ nodes.
The degree of each node is equal to two. 

\begin{figure}
\centering
\includegraphics[width=0.45\hsize]{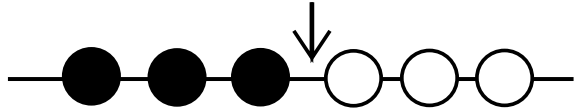}
\caption{
Schematic of interface on the ring.
The local network surrounding an interface (pointed by the arrow) is depicted.
Open and solid circles represent voters with opinions {\bf 0} and {\bf 1}, respectively. 
}
\label{fig:interface-schematic}
\end{figure}

\begin{figure}
\centering
\includegraphics[width=0.45\hsize]{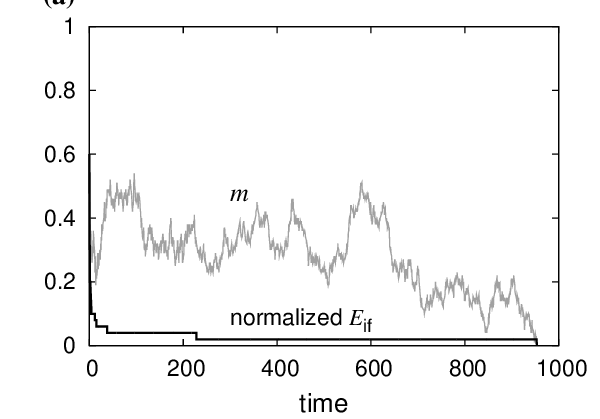}
\hspace*{3mm}
\includegraphics[width=0.45\hsize]{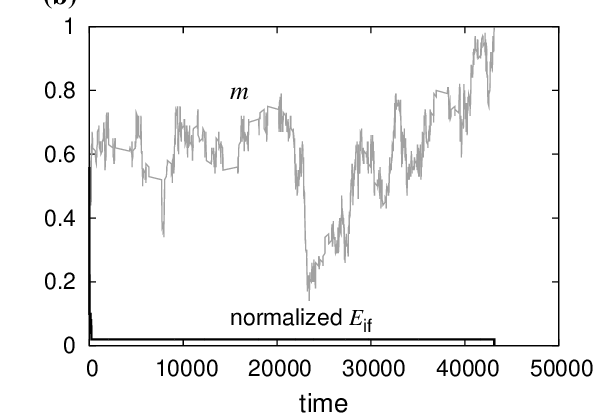}
\caption{
Time course of the fraction of  {\bf 1} voters $m$ (gray lines) and the number of
interfaces $\Eif$ (black lines) in the (a) exponential and (b) power-law voter models
on the ring with 100 nodes.
The values of $\Eif$ shown are those normalized by the total number of links in the ring, which is equal to $N$.
}
\label{fig:m-ai-ring}
\end{figure}

To understand the mechanism governing the consensus time, denoted by $T$,
we begin with tracking the fraction of voters who take opinion {\bf 1} and the number of the interfaces,
denoted by $m$ and $\Eif$, respectively.
For the ring, a link is defined to be an interface 
when the two endpoints of the link are occupied by the opposite opinions (\FIG\ref{fig:interface-schematic}).
In the case of the ring, an interface separates a domain of {\bf 0}s and a domain of {\bf 1}s.
Figures~\ref{fig:m-ai-ring}(a) and \ref{fig:m-ai-ring}(b) represent an example time course of $m$ and $\Eif$
for the exponential and power-law voter models on the ring with $N = 100$, respectively.
Because two interfaces that meet on a link annihilate each other and decrease $\Eif$ by two
and because new interfaces are not produced inside a domain containing a single opinion,
$\Eif$ monotonically decreases for both the models.
Figure~\ref{fig:m-ai-ring} also indicates that $\Eif=2$ for most of the time before consensus. 
When $\Eif=2$,
the ring consists of one domain of {\bf 0}s and one domain of {\bf 1}s.
Therefore, $T$ can be approximated by the time at which the two interfaces collide
since $\Eif$ decreases to two.
For the exponential voter model on the ring,
the distance between the two interfaces follows the simple random walk.
Therefore, $T$ is estimated to be the time needed for the random walker
to travel a distance of ${\rm O}(N)$; \ie, $T \propto N^2$~\cite{Castellano2009spo,Cox1989crw}.
Because the probability with which an interface moves to the left and that to the right are equal to $1/2$
independent of the detail of $p(\tau)$,
the number of the movements of the interface until the consensus is reached for the power-law voter model
is the same as that of the exponential voter model.
Therefore, $T$ for the power-law voter model also obeys $T \propto N^2$.

Therefore,
the effect of the power-law $p(\tau)$ on $T$ for the ring
can be ascribed to the difference in the behavior of the interval between successive movements of an interface.
We refer to this interval as the sojourn time of the interface and denote it as $s$.
For the exponential and power-law voter models on the ring,
$T$ is roughly proportional to the product of $\langle s \rangle$ and $N^2$. 
Therefore, the ratio of $T$ for the power-law voter model to that of the exponential voter model is determined by $\langle s \rangle$.

\begin{figure}
\centering
\includegraphics[width=0.45\hsize]{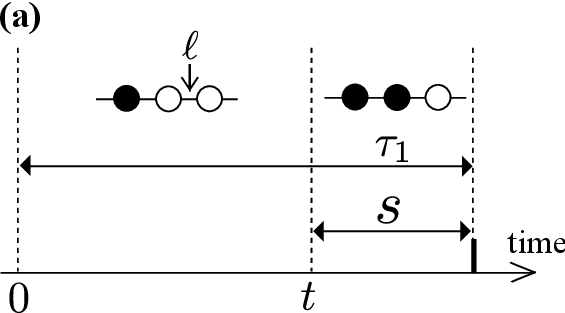}
\hspace*{3mm}
\includegraphics[width=0.45\hsize]{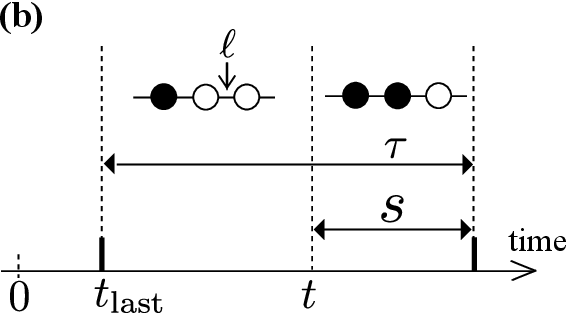}
\caption{
Schematics of the movement of an interface on the ring.
In (a), no update event occurs on link $\ell$ between time $0$ and time $t$.
In (b), at least one update event occurs between time $0$ and time $t$.
In (b), the time of the last update event before $t$ is denoted as $t_{\rm last}$.
The nodes and links near link $\ell$ on the ring are depicted above the time lines of the update events. 
Open and solid circles represent voters with opinion {\bf 0} and {\bf 1}, respectively. 
}
\label{fig:ring-schematic}
\end{figure}

To derive $\langle s \rangle$ for the exponential and power-law voter models,
we consider the situations illustrated in \FIGS\ref{fig:ring-schematic}(a) and \ref{fig:ring-schematic}(b).
The time lines in the figures indicate the renewal process on link $\ell$.
We refer to the situations as case~(a) and case~(b), respectively.  
Suppose that an interface moves to the right at time $t$.
Link $\ell$ shown in \FIG\ref{fig:ring-schematic} becomes an interface at time $t$
owing to the occurrence of an event on the adjacent link.
In case~(a), no update event is assumed to have occurred on link $\ell$ between time $0$ and time $t$.
Time $t$ is included in IEI $\tau_1$, which obeys $p_1(\tau_1)$.
Therefore, $s = \tau_1 - t$ in case~(a).
In case~(b), at least one update event is assumed to have occurred on link $\ell$ between time $0$ and time $t$.
We denote the time at which the last update event occurs on $\ell$ before time $t$ as $t_{{\rm last}}$.
Time $t$ is included in IEI $\tau$ that was drawn from distribution $p(\tau)$ at time $t_{\rm last}$.
Therefore, $s = \tau - (t-t_{\rm last})$ in case~(b).

We calculate $\langle s \rangle$ of the exponential and power-law voter models
for cases~(a) and (b).
In case~(a),
the probability that $\tau_1 > \tau^\prime$ conditioned by $t$ is given by
\begin{eqnarray}
{\rm Pr}(\tau_1 > \tau^\prime \mid t)
&=\frac{\int_{\tau^\prime}^\infty p_1(u^\prime) du^\prime}{\int_{t}^\infty p_1(u^\prime) du^\prime}
&=\frac{P_1(\tau^\prime)}{P_1(t)},
\end{eqnarray}
where
\begin{equation}
P_1(u) = \int_{u}^\infty p_1(u^\prime)du^\prime.
\end{equation}
Therefore, we obtain
\begin{eqnarray}
p_1(\tau_1 \mid t) 
&=\frac{d}{d\tau_1}\left[ 1 - \frac{P_1(\tau_1)}{P_1(t)} \right]
&=\frac{p_1(\tau_1)}{P_1(t)}.
\label{eq:p1(tau1)}
\end{eqnarray}
Because $\tau_1 = s + t$ (see \FIG\ref{fig:ring-schematic}(a)),
we obtain
\begin{equation}
p_s(s | t) = \frac{p_1(s + t)}{P_1(t)}.
\label{eq:pw1}
\end{equation}

For the exponential voter model,
the substitution $p_1(\tau_1)=\lambda \exp(-\lambda \tau_1)$ in \EQ\eqref{eq:pw1} yields
\begin{equation}
p_s(s| t) = \lambda \exp(-\lambda s).
\label{eq:pw1-exp}
\end{equation}
Therefore, $s$ obeys the same exponential distribution as that obeyed by $\tau$, and we obtain
\begin{equation}
\langle s \rangle^{\exp} = \langle \tau \rangle,
\label{eq:tauw1-exp}
\end{equation}
where $\langle s \rangle$ denotes the mean of $s$
with respect to the density $p_s(s | t)$.
For the power-law voter model,
the substitution $p_1(\tau_1) = \left[ \left( \tau_1+c \right) / c \right]^{-(\alpha-1)}$ in \EQ\eqref{eq:pw1} yields
\begin{equation}
p_s(s | t) = \frac{\alpha-2}{t+c} \left[ \frac{s +t+c}{t+c} \right]^{-(\alpha-1)}.
\end{equation}
This conditional density is of the same form as $p_1(\tau_1)$ for the power-law voter model
given by \EQ\eqref{eq:p1_power},
with $c$ in \EQ\eqref{eq:p1_power} replaced by $t + c$ and the exponent $\alpha-1$ unchanged.
Therefore, we obtain
\begin{equation}
\langle s \rangle^{\rm power} =
\left\{
\begin{array}{ll}
+\infty & (\alpha \leq 3),\\
\frac{\alpha-2}{\alpha-3} \left( \langle \tau \rangle + \frac{t}{\alpha-2} \right) & (3 < \alpha).
\end{array}
\right.
\label{eq:tauw1-EP}
\end{equation}

On the basis of \EQS\eqref{eq:tauw1-exp} and \eqref{eq:tauw1-EP}, we obtain
\begin{equation}
\langle s \rangle^{\exp} < \langle s \rangle^{\rm power}
\label{eq:<tauw>-a} 
\end{equation}
in case~(a), regardless of the value of $\alpha$.

In case~(b),
the probability density of $s$ conditioned by $t-t_{\rm last}$ is given by 
\begin{equation}
p_s(s | t-t_{\rm last}) =
\frac{p(s + t-t_{\rm last})}{P(t-t_{\rm last})},
\label{eq:pw}
\end{equation}
through a derivation similar to that of \EQ\eqref{eq:pw1},
where
\begin{equation}
P(u) = \int_{u}^\infty p(u^\prime)du^\prime.
\end{equation}

For the exponential voter model,
the substitution $p(\tau)=\lambda \exp(-\lambda \tau)$ in \EQ\eqref{eq:pw} yields
\begin{equation}
p_s(s | t-t_{\rm last}) = \lambda \exp(-\lambda s),
\end{equation}
that is, $s$ is statistically the same as in case~(a) (\EQ\eqref{eq:pw1-exp})
and $s$ obeys the same exponential distribution as that obeyed by $\tau$.
Therefore, we obtain
\begin{equation}
\langle s \rangle^{\exp} = \langle \tau \rangle,
\label{eq:tauw-exp}
\end{equation}
where $\langle s \rangle$ denotes the mean of $s$
with respect to the density $p_s(s | t-t_{\rm last})$.
For the power-law voter model,
the substitution $p(\tau) = \left[ \left(\alpha-1 \right)/c \right] \left[ \left( \tau+c \right) / c \right]^{-\alpha}$
in \EQ\eqref{eq:pw} yields
\begin{equation}
p_s(s | t-t_{\rm last}) =
\frac{\alpha - 1}{t-t_{\rm last} + c}\left[\frac{s + t-t_{\rm last} + c }{t-t_{\rm last} + c} \right]^{-\alpha}.
\label{eq:pw-EPOP}
\end{equation}
Equation~\eqref{eq:pw-EPOP} is of the same form as \EQ\eqref{eq:p_power},
with $c$ in \EQ\eqref{eq:p_power} replaced by $t-t_{\rm last} + c$ and $\alpha$ unchanged.
Therefore, we obtain
\begin{equation}
\langle s \rangle^{\rm power} =
\langle \tau \rangle +\frac{t-t_{\rm last}}{\alpha-2}.
\label{eq:tauw-EPOP}
\end{equation}
Although $\langle s \rangle^{\rm power}$ increases with $t-t_{\rm last}$,
it does not diverge because $t-t_{\rm last}$ on link $\ell$ is finite by definition.

On the basis of \EQS\eqref{eq:tauw-exp} and \eqref{eq:tauw-EPOP}, we obtain
\begin{equation}
\langle s \rangle^{\exp} < \langle s \rangle^{\rm power}
\label{eq:<tauw>-b} 
\end{equation}
in case~(b).

\begin{figure}
\centering
\includegraphics[width=0.45\hsize]{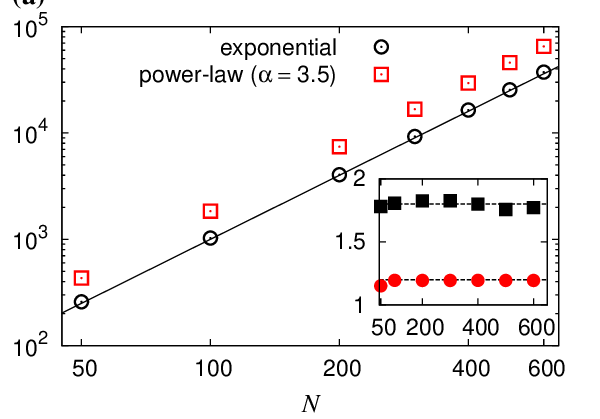}
\includegraphics[width=0.45\hsize]{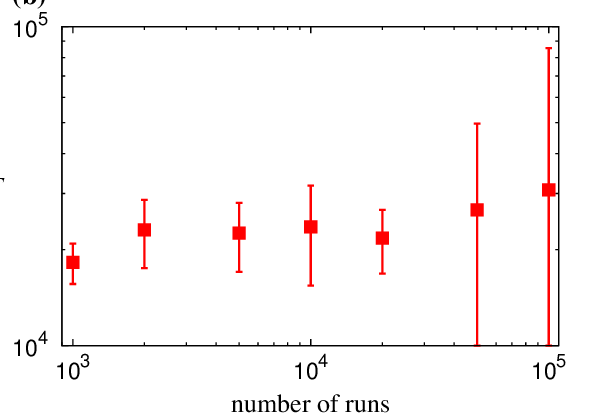}
\caption{(Color online) (a) Average consensus time of the exponential voter model (circles)
and power-law voter model with $\alpha=3.5$ (squares) on the ring (main panel).
The solid line indicates $T \propto N^2$.
For the results shown in the main panel,
$T^{\rm power} / T^{\exp}$ (squares)
and $\langle t - t_{\rm last} \rangle$ for the power-law voter model (circles) are plotted against $N$ in the inset.
(b)  Average consensus time of the power-law voter model with $\alpha=2.5$
on the ring as a function of the number of simulation runs. 
The error bars indicate the standard deviation.
We set $N=100$.
}
\label{fig:avgconstime-ring}
\end{figure}

Equations~\eqref{eq:<tauw>-a} and \eqref{eq:<tauw>-b}
predict that $T$ for the power-law voter model is larger than that for the exponential voter model.
The mechanism governing the enlarged consensus time for the power-law voter model is essentially the same as
that governing the slowing down of epidemic dynamics in the case of long-tail IEI distributions~\cite{AVazquez2007ion,Iribarren2009ioh,Min2010sdf,Karsai2010sbs}.
$T$ for the exponential voter model and power-law voter model with $\alpha=3.5$ on the ring
are shown in \FIG\ref{fig:avgconstime-ring}(a).
The numerical results are consistent with the theoretical prediction $T^{\rm exp} < T^{\rm power}$.
We find $T^{\rm power} / T^{\rm exp} \approx 1.8$ for different values of $N$
(inset of \FIG\ref{fig:avgconstime-ring}(a)).
To understand the origin of the value of $T^{\rm power} / T^{\rm exp}$,
we note that $T^{\rm power} / T^{\rm exp} \approx\langle s \rangle^{\rm power} / \langle s \rangle^{\exp}$
because most of the movements of the interfaces correspond to case~(b) when $t$ is sufficiently large.
We approximate $t-t_{\rm last}$ in $\langle s \rangle^{\rm power}$ (\EQ\eqref{eq:tauw-EPOP}) by $\langle t-t_{\rm last} \rangle \approx 1.2$, which we obtained from the direct numerical simulations of the power-law voter model on the ring. 
Then, the substitution $\langle \tau \rangle = 1$ and $\alpha=3.5$ in \EQS\eqref{eq:tauw-exp} and \eqref{eq:tauw-EPOP}
yields $T^{\rm power} / T^{\rm exp} \approx 1.8$.

Equation~\eqref{eq:tauw1-EP} indicates that
$T$ diverges for the power-law voter model with $\alpha \leq 3$.
In numerical simulations with $\alpha = 2.5$,
we confirmed that $T$ increases with the number of runs, albeit slowly~(\FIG\ref{fig:avgconstime-ring}(b)).

\subsection{Complete graph}\label{sec:completegraph}
\begin{figure}
\centering
\includegraphics[width=0.6\hsize]{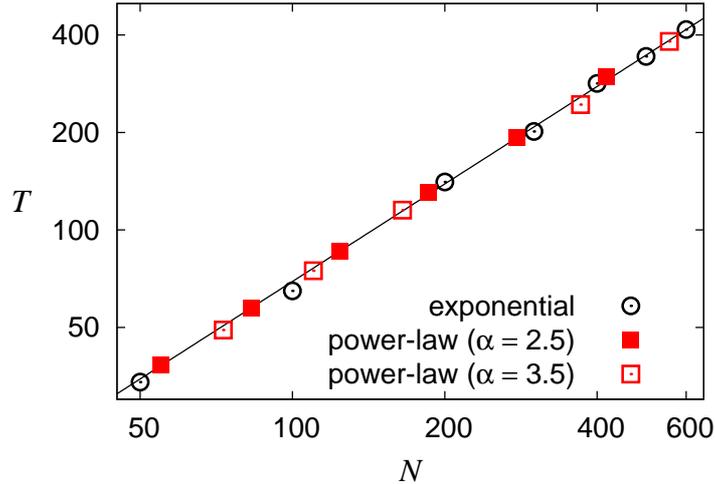}
\caption{(Color online) Average consensus time of the exponential (circles) and power-law (squares) voter models on the complete graph with $N$ nodes.
The solid line is the analytical solution given by \EQ\eqref{eq:T_complete} for the exponential voter model;
\ie, $T = N \ln 2$.}
\label{fig:avgconstime-all}
\end{figure}

In this section, we examine the exponential and power-law voter models on the complete graph. 
The average consensus time $T$ for the exponential voter model is given by~\cite{Sood2008vmo}:
\begin{equation}
T = N
\left[ m_0 \log \frac{1}{m_0} + (1 - m_0) \log \frac{1}{1 - m_0} \right],
\label{eq:T_complete}
\end{equation}
where $m_0$ denotes the initial value of $m$.
Figure~\ref{fig:avgconstime-all} shows the plot of $T$ for the exponential and power-law voter models on the complete graph.
The value of $T$ for the power-law voter model with $\alpha=2.5$ and $\alpha=3.5$ are close to that for the exponential voter model.

In contrast to the case of the ring, 
we cannot understand the behavior of $T$ for the two voter models
on the complete graph by tracking the position of a single interface.
Therefore, we focus on the sequences of the effective events.
We define the effective event as the update event that occurs on an inconsistent link
whose two endpoints have the opposite opinions (essentially the same as \FIG\ref{fig:interface-schematic}).
If an update event is an effective event, magnetization $m$ changes by $1/N$.
Otherwise, the update event does not affect $m$.
We denote the interval between successive effective events on the network as $\taue$.
$T$ is equal to the sum of $\taue$ until the consensus. 
Note that $\taue$ is generally large when there are relatively few inconsistent links.

\begin{figure}
\includegraphics[width=0.45\hsize]{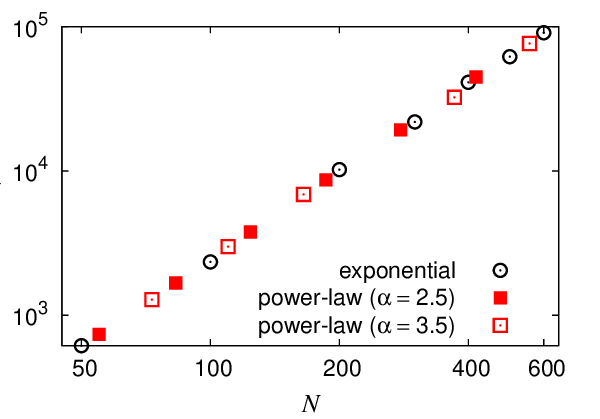}
\hspace*{3mm}
\includegraphics[width=0.45\hsize]{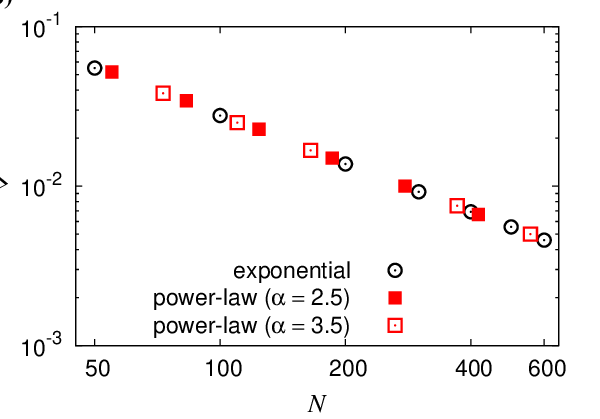}
\caption{(Color online) 
(a) Average number of the effective events $\langle \Me \rangle$ 
and (b) average interval between successive effective events $\langle \taue \rangle$
on the complete graph
for the exponential (circles) and power-law (squares) voter models.}
\label{fig:effectupdate-all}
\end{figure}

In \FIG\ref{fig:effectupdate-all}, we plot the values of  $\langle \Me \rangle$ and $\langle \taue \rangle$
until the consensus for the two voter models. 
We define $\langle \Me \rangle$ as the average number of the effective events until the consensus.
$\langle \Me \rangle$ is almost the same for the exponential and power-law models and scales as $N^2$, as shown in \FIG\ref{fig:effectupdate-all}(a).
$\langle \taue \rangle$ is almost the same for the two voter models and scales as $N^{-1}$, as shown in \FIG\ref{fig:effectupdate-all}(b).
Figure \ref{fig:effectupdate-all} is consistent with \FIG\ref{fig:avgconstime-all};
$T$ is equal to the product of $\langle \Me \rangle$ and $\langle \taue \rangle$.

In the rest of this section, we examine $\langle \taue \rangle$ for the two voter models.
Suppose that an effective event occurs at time $t$.
The time to the next effective event $\taue$ is equal to
the smallest sojourn time among those of the $E_{\rm i}$ inconsistent links.
We estimate $\taue$ by the extremal criterion
\begin{equation}
\int_0^{\taue} p_{\rm i}(\taui) d\taui \approx \frac{1}{\Ei},
\label{eq:minimum_taue}
\end{equation}
where $p_{\rm i}(\taui)$ represents the probability density of 
forward recurrence time $\taui$ of an inconsistent link 
and is given by~\cite{Cox1967rt}:
\begin{equation}
p_{\rm i}(\taui) = \frac{1}{\langle \tau \rangle}P(\taui).
\end{equation}

For the exponential voter model, by substituting
\begin{equation}
p_{{\rm i}}(\taui) = \lambda\exp(-\lambda\taui)
\label{eq:pw-exp}
\end{equation}
in \EQ\eqref{eq:minimum_taue}, we obtain
\begin{equation}
\taue^{\exp} = -\frac{1}{\lambda}\log \left( 1-\frac{1}{\Ei} \right).
\label{eq:taue-exp}
\end{equation}
For the power-law voter model, by substituting 
\begin{equation}
p_{{\rm i}}(\taui) = \left( \frac{\taui + c}{c} \right)^{-(\alpha-1)}
\label{eq:pw-EP}
\end{equation}
in \EQ\eqref{eq:minimum_taue}, we obtain
\begin{equation}
\taue^{{\rm power}}
= c \cdot \left[ -1 + \left( 1 - \frac{1}{\Ei} \right)^{-\frac{1}{\alpha-2}} \right]. 
\label{eq:taue-EP}
\end{equation}

\begin{figure}
\centering
\includegraphics[width=0.6\hsize]{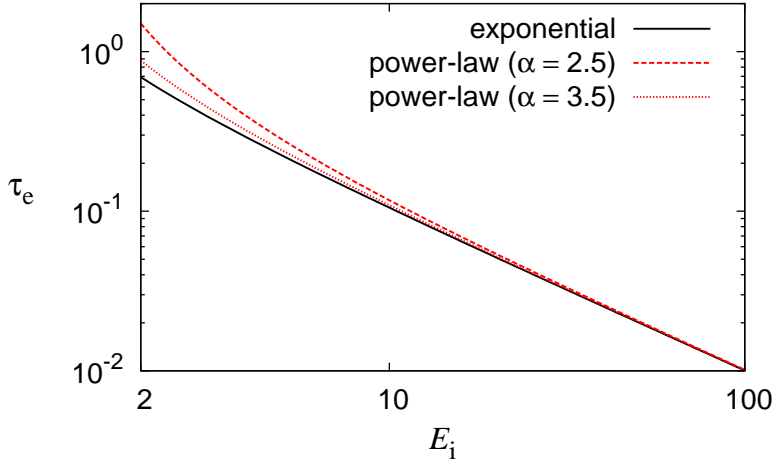}
\caption{(Color online) Expected interval between effective update events $\taue$
for the exponential and power-law voter models, \ie, \EQS\eqref{eq:taue-exp} and \eqref{eq:taue-EP}.}
\label{fig:taue-exp-EP}
\end{figure}

In \FIG\ref{fig:taue-exp-EP}, $\taue^{\exp}$ and $\taue^{\rm power}$ are plotted against $\Ei$.
For the complete graph, $\Ei = m(1-m)N^2$ is relatively large as compared with that for the ring.
For a sufficiently large $\Ei$, both \EQS\eqref{eq:taue-exp} and \eqref{eq:taue-EP} are approximated by $\langle \tau \rangle / \Ei$.
Therefore, we obtain $\taue^{\exp} \approx \taue^{\rm power}$ for a large $\Ei$,
which is consistent with the results shown in \FIG\ref{fig:taue-exp-EP}
and the numerically obtained $\langle \taue \rangle$ shown in \FIG\ref{fig:effectupdate-all}(b).
The results shown in \FIG\ref{fig:taue-exp-EP} are also consistent with the results for the ring, 
for which $\Eif = \Ei = 2$ for most of the time (\FIG\ref{fig:m-ai-ring}).
As shown in \FIG\ref{fig:taue-exp-EP}, $\taue^{\rm power}$ is considerably larger than $\taue^{\exp}$ for a small $\Ei$.
Therefore, $T \propto \langle \taue \rangle N^2$ is presumably larger
for the power-law voter model than for the exponential voter model on the ring,
which is actually the case (\FIG\ref{fig:avgconstime-ring}).

\subsection{Extended rings and regular random graphs}
$T^{\rm power} / T^{\exp}$ is different for the ring (\SEC\ref{sec:ring}) and the complete graph (\SEC\ref{sec:completegraph}).
Therefore, the effect of the power-law IEIs on $T^{\rm power} / T^{\exp}$ may depend on the degree of the node. 

\begin{figure}
\includegraphics[width=0.45\hsize]{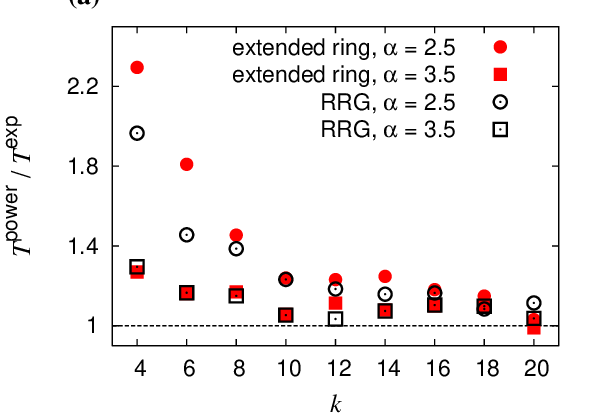}
\hspace*{3mm}
\includegraphics[width=0.45\hsize]{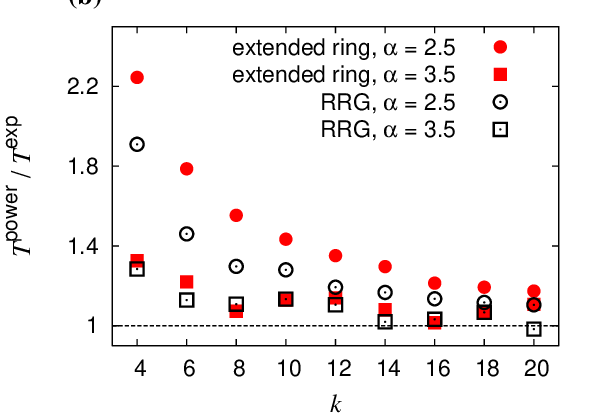}
\caption{(Color online) $T^{\rm power} / T^{\exp}$ for the extended ring (filled symbols) and for the RRG (empty symbols) with degree $k$.
We set (a) $N=500$ and (b) $N=1000$.}
\label{fig:rT_k}
\end{figure}

To show this, we first investigate the consensus time on the extended ring with degree $k$,
where $k$ is an even number and each node is connected to up to the $(k/2)$th nearest neighbors
on each side of the conventional ring. 
The extended ring with $k=2$ is equivalent to the conventional ring considered in \SEC\ref{sec:ring}.
The extended ring approaches the complete graph as $k \to N-1$. 
For the extended ring, $T^{\rm power} / T^{\exp}$ is shown as a function of $k$ by filled symbols in \FIGS\ref{fig:rT_k}(a) and \ref{fig:rT_k}(b)
for $N=500$ and $N=1000$, respectively.
$T^{\rm power} / T^{\exp}$ decreases and approaches unity as $k$ increases.

To further examine if $T^{\rm power} / T^{\exp}$ decreases with $k$,
we investigate the consensus time for the regular random graph (RRG) with degree $k$.
We generate the RRG by using the configuration model~\cite{Newman2001rgw} as follows.
Each node is initially given $k$ stubs, \ie, half links.
Then, two stubs are chosen randomly with equal probability.
We connect the two stubs to create a link
unless a self-loop or multiple links are generated;
in such a case, we discard the selected pair of stubs.
We repeat this procedure until all the stubs are exhausted
to obtain an instance of the RRG with degree $k$. 
If the procedure is stuck midway or the generated network is not connected,
we restart the procedure.

For the RRG, $T^{\rm power} / T^{\exp}$ is shown as a function of $k$ by empty symbols in \FIGS\ref{fig:rT_k}(a) and \ref{fig:rT_k}(b) for $N=500$ and $N=1000$, respectively.
$T^{\rm power} / T^{\exp}$ decreases and approaches unity as $k$ increases, which is qualitatively the same as the results for the extended ring.

\section{Conclusions}
\subsection{Summary of the results}
In this study, we numerically investigated voter models with the power-law IEI distribution.
On the ring, the consensus time for the power-law voter models is larger than
that for the exponential (\ie, standard) voter model.
On the complete graph, the ratio of the consensus time for the power-law voter model to that for the exponential voter model
becomes relatively small compared to the ratio in the case of the ring.
We numerically investigated the two voter models on the extended ring and the regular random graph,
and we confirmed that the effect of the power-law IEI distribution to enlarge the consensus time decreases with the degree of the node.
The difference in the consensus time originates from the fact that
the expected time until the next update event on a link for the power-law voter model 
is large if the link has not had an update event for a long time.
On the ring, the interval between successive movements of the interface (\ie, the link connecting nodes with the opposite opinions)
is elongated by the memory effect for the power-law distribution.
On the complete graph, the dynamics is determined by the smallest waiting time among many inconsistent links,
which can be nearly identical for the exponential and power-law distributions. 
 
\subsection{Relation to previous work}
Long-tail IEI distributions
are known to make epidemic dynamics slower~\cite{AVazquez2007ion,Iribarren2009ioh,Karsai2010sbs,Min2010sdf}.
Our results for the ring are consistent with these results;
the occurrence of extremely large IEIs at some links crucially slows down the dynamics.

Antal and colleagues considered three update rules, \ie,
the link dynamics (LD), the voter model (VM) and the invasion process (IP)~\cite{Sood2008vmo,Antal2006edo} (also see \cite{Ohtsuki2006asr,Suchecki2005clf,FVazquez2008aso}).
The consensus time for the LD, VM, and IP is considerably different on heterogeneous networks~\cite{Sood2008vmo,Antal2006edo}.
We adopted the LD update rule in this study.
We can define the VM and the IP on regular graphs with degree $k$ for a general
IEI distribution $p(\tau)$ as follows.
Initially, each voter, not each link, is independently assigned
with a random IEI until the initial update event occurs according to the distribution $p_1(\tau_1)$.
Suppose that an update event occurs at a voter.
In the VM, the voter adopts the opinion of a neighbor
that is selected with the equal probability $1/k$ from the neighborhood.
In the IP, the voter imposes its opinion on a neighbor that is selected with probability $1/k$.
For either update rule, the next IEI for the voter is drawn from $p(\tau)$.
On the complete graph with $N=200$, $\lambda = 1$, and $(\alpha,c)=(2.5,0.5)$,
for example, we obtain $T \approx 1.35 \times 10^2$ and $3.84 \times 10^4$
for the exponential and power-law voter models with the VM update rule, respectively.
With the VM, $T$ for the power-law voter model is larger than that for the exponential voter model.
This result is in contrast to that with LD; 
$T$ with the LD is almost the same between the exponential and power-law voter models (\SEC\ref{sec:completegraph}).
We note that even on regular graphs,
$T$ with the power-law IEI distribution depends on the update rule,
where the LD, VM, and IP are equivalent in the case of the exponential IEI distribution~\cite{Sood2008vmo,Antal2006edo}.

Finally,
we remark on the relationship between our model and two other models that were recently proposed.
Stark and colleagues examined a variant of the voter model with the VM
update rule and an increasing inertia of voters~\cite{Stark2008dmc}.
The inertia implies that the transition rate
at which a focal voter imitates the opinion of a neighbor
decreases with the time since the latest change in the focal voter's opinion.
They found that an appropriate amount of inertia
shortens the consensus time on several networks.
This result is opposite to our preliminary results for the VM described above.
Although the reason of this inconsistency is not clear,
we point out two differences in the two models.
First, the transition rate can be infinitesimally small in the VM with the power-law IEI distribution,
whereas it has a lower bound in Stark's model. 
Second, when a voter experiences an update event that does not change its opinion,
the transition rate of the voter is reset in the VM with the power-law IEI distribution,
whereas the transition rate is not affected by such an update event in Stark's model.

Fern\'{a}ndez-Gracia and colleagues investigated other variants of power-law voter models~\cite{Fernandez-Gracia2011}.
One of their models in which the rate at which a voter experiences an update event decreases with the time since the last update event of the voter  (called exogenous update in \cite{Fernandez-Gracia2011}) is equivalent to the VM variant of our power-law voter model.
They also considered the update rule in which update events occur at a constant rate
(called random asynchronous update in \cite{Fernandez-Gracia2011}),
which is equivalent to the VM variant of our exponential voter model.
For the complete graph, they found that $T$ with the exogenous update is larger than that with the random asynchronous update.
This result is consistent with our preliminary result  described above.

\section*{Acknowledgments}@
This work is supported by Grant-in-Aid for JSPS Fellows from Japan Society for the
Promotion of Science (JSPS).
N.M. acknowledges the support provided through Grants-in-Aid
for Scientific Research (No. 20760258 and No. 23681033) from MEXT, Japan.


\end{document}